# EJECTA SPEED WITH THE PROJECTILE VELOCITY LESS THAN 2 KM/S

Hector J. Durand-Manterola[1,2], Alvaro Suárez-Cortés[3]

1 Corresponding author durand_manterola@igeofisica.unam.mx
2 Department of Space Sciences, Geophysics Institute, National Autonomous University of Mexico, Mexico City, Mexico.
3 Postgraduate in Earth Sciences, National Autonomous University of Mexico, Mexico City, Mexico.

ABSTRACT
This work is a theoretical study of the speed at which the material of an impacted target is ejected during the formation of an impact crater. Our model, starting from the first principle of thermodynamics, can describes the speed of the ejecta recursing to considerations that include complex process in simple calculations.
The fit of the model with observations shows that the many complex details implicit in an impact process could be included in some few parameters (D, $v_0$, $r_0$, $\Delta r$). Ejecta speed could be described independent of impactor parameters.
The model is compared with subsonic and supersonic speed experiments showing coincidence in several cases. The model works with subsonic and supersonic impacts.
We do not compare the model with hypersonic impacts (> 5 km/s), however, as the model derivation no depends on the impactor velocity it is likely that also work with this kind of impacts.

1 INTRODUCTION
Impact of meteoroids, asteroids, or comets on a solid planetary surface produces, the most of times, circular structures called impact craters. Occurrence of the impacts is random; arrival of an impactor cannot be predicted regularly since its distribution in space is random. Crater formation occurs when kinetic energy of the impactor (asteroid, comet or meteoroid) is transmitted to the target (impacted body surface). This produces a compression wave or a shock wave (hereinafter only mentioned as "the wave"), depending on the impactor speed. The wave propagates through the target and the pressure increase breaks up the target material and throws it at a certain speed, excavating a crater and depositing the thrown material around it. The term "ejecta" refers to the material ejected during an impact crater formation. Solid ejecta is usually radially distributed from the crater rim above the planetary surface.
When an impactor hits a planetary body, in all cases the speed of the impactor exceeds the escape velocity of the planet. However, in its great majority, ejecta reaches lower speeds than the impactor, depositing on the surface of the body. Only a very small fraction reaches escape velocity, and this is the origin of the Martian and Lunar meteorites.
The formation of an impact crater is extremely complex phenomena that involve dynamic, mechanical, thermodynamic and geochemical processes (Melosh (1989) and French (1998)). The variables involved in planetary-scale impacts are diverse and in most cases are unknown (Melosh (2011)), for this reason it has been resorted to simulating the processes of impact crater formation through experimentation in laboratories, impacting various projectiles on various surfaces at speeds that have escalated from low speeds (<1 km / s) to increasing speeds (> 1 km / s) (Housen and Holsapple (2011)).

In the impact of a projectile on a target and in which a crater is generated, the dispersed material -essentially of the target- form a transitory conical-type structure that surrounds the impact zone where most of the ejected particles follow ballistic trajectories (the envelope of this structure is called the ejection curtain (Melosh (1989)), its dynamic and morphological characteristics depend, presumably, on the dynamic and mechanical properties of the projectile and on the mechanical properties of the target or impact surface, also on the structure of both. Melosh (2011) and French (1998) divide the phenomenon of impact crater formation into three stages: contact and compression, excavation and modification. Regarding this document, we will focus excavation stage

In the contact \ compression stage, it begins at the instant the projectile touches the surface and ends with the transfer of its energy (disintegration of the projectile). The transferred energy is quickly translated into shock waves that propagate spherically at high speeds through the two interacting bodies. The dissipation of impact energy is translated into energy in the form of heat and through deformation and acceleration of the material that will later be excavated to form the crater (French, 1998). In the excavation stage the impact crater begins to take shape, through complex interactions between the expanding shock waves and the original soil surface. Some of the energy from the initial shock wave is converted back to kinetic energy displacing material outward, largely as individual fragments traveling at high speeds (French, 1998). This process continues until the shock and release waves are no longer able to excavate any more material from the surface.

The study of the speed that ejecta has when it is expelled has been approached from two angles: First, experimental studies in which an impact is produced in the laboratory and the velocities of the ejecta are measured as a function of the distance to the impact point; second, studies from a point of view of the dimensional analysis, looking for a scaling of these speeds with the speed of the impactor; and in some cases, hybrid studies with the two approaches (Hartman, 1985); Anderson et al., 2003; Michikami et al., 2007; Hermalyn and Schutz, 2010).

In this work we develop a mathematical model deduced from first principles, explicitly, from first thermodynamics law.

The ejecta speed obtained is dependent of distance from impact point and some parameters of the wave. As **in the model** the ejecta speed does not depends on the impactor velocity, it is expected that the model is suitable in all range of impact velocities. However, in this work the model it is only compared with subsonic and supersonic experiments (10 m/s < v < 2 km/s) and not with ultrasonic impacts because we do not find ejection velocity measures in literature**.** Comparison of the model with experimental data, shows an acceptable fit.

In section 2 the model it is described; in section 3 the model it is compared with experimental data; in section 4 it is discussed the results, and in section 5 the conclusions are presented. The formal development of the model, in form of six theorems are in the appendix.

2. MATHEMATICAL MODEL
2.1 General Hypotheses
To build our model we assume the next hypotheses: our system of study is **a** solid hemisphere of the target that is removed to form the crater (see Figure 1). The point of impact, O, is in the center of the sphere. The radius of the sphere is R; it is equal to the transient crater radius. The density of **t**he system material, $\rho$, is constant and equal in all points. The energy that the impactor carry is only kinetic, and at the contact with the system in point O, it is all transferred to the target. A wave propagates radially from point O in all directions inside the hemisphere. In our model we consider the impact process beginning at the contact of the projectile with the target until last ejecta was expel.

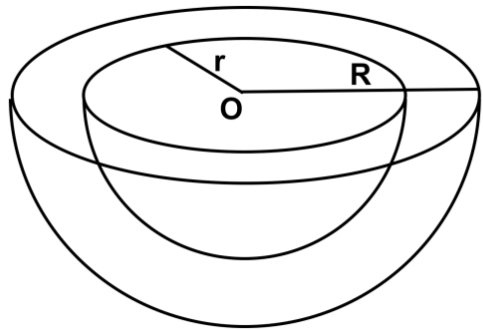

**Figure 1. Diagram of studied system**

2.2 System's Energy
An impact process, i.e., the contact of a projectile with a target, and the excavation of the crater, is essentially a simple phenomenon, the transference of energy from the projectile to the target. In all the process E (the total energy transfer from the projectile) is constant because the energy of the projectile in the moment of contact is constant. However, the way where kinetic energy of the projectile transforms in kinetic energy of the ejecta is not immediate. As it is said: the devil is in the details. From a thermodynamic point of view, we can simplify calculation, distributing target's energy in three parts and starting with the first law**,** as follows:

$$E = E_1 + E_2 + E_3 \quad (1)$$

This could seen a very simplified approach of a more complex process, but the three divisions considered include several subprocess. $E_1$ include only the ejecta movement that is the main objective of the model. $E_2$ includes, in part energy for fragmentation, in part for compress the material, in part as propagation energy of the wave. $E_3$ energy includes heating, melting, vaporization, frictional losses and the retained kinetic energy by the projectile and the target. By different process of dissipation this energy finally converts in thermal energy, i.e., internal energy of the target material.
The energy E₂, carried by the wave is:

$$E_2 = PV_W \quad (2)$$

Where P is the mean pressure into the wave, and $V_W$ is the volume occupied by the wave. This is the energy contained in the wave, it is not an expansion work.

As the wave pass, the material of the target is removed, then behind the wave do not exist a rarefaction wave that carry energy. In all moment, during the excavation, the wave is situated in the boundary of excavation. The concept of rarefaction wave is true if the material remains in its place but in an impact process the material is removed leaving only vacuum. In addition, most of the energy of a pressure wave is in the zone of compresion (pressure is energy per unit volume).

We assume that the impactor energy mainly spends in the processes of fragmentation, acceleration of the material (i.e., transformation in kinetic energy), and propagation of the wave, the other processes spend a lesser amount of energy and it can be neglected. Then we assume:

$$E_3 = 0 \quad (3)$$

Let H be the fraction of the system of radius r (see figure 1), excavated or removed by the impact at some moment, and let v be the speed of the ejected material at the moment of leave the ground. Kinetic energy, $E_1$, of H is (see appendix, theorem 1):

$$E_1 = \rho\pi \int_0^r v^2 r^2 dr \quad (4)$$

Using equations (1), (2), (3) and (4) the energy E is expressed as follows:

$$E = \rho\pi \int_0^r v^2 r^2 dr + PV_W \quad (5)$$

The wave propagation energy by unit volume is the pressure inside the wave. This pressure, being greater than tensile strength in target material, is the responsible of its fragmentation and the acceleration of ejecta. From theorem 3 (see appendix) we obtain an expression of the pressure:

$$P = Dg\frac{1}{2}\rho v^2 \quad (6)$$

D is constant in each event but variable from event to event. The value of D depends, implicitly, for cohesion or yield strength of the material.
As Dg > 1 (see appendix, theorem 4), then pressure is greater than kinetic energy density (theorem 2):

$$P > \frac{1}{2}\rho v^2 \quad (7)$$

An expression of $V_W$ is:

$$V_W = 2\pi r^2 \Delta r \quad (8)$$

Substituting equation (6) and (8) into equation (5) we obtain:

$$E = \rho\pi \int_0^r v^2 r^2 dr + Dg\pi\rho v^2 r^2 \Delta r \quad (9)$$

2.3 Speed of the Ejecta

This model only takes in count the velocity of the ejecta in the moment of be launched from the ground and do not examine the trajectory followed underground and later at the ballistic flight.

The rock of the target is fragmented (when the target is a cohesive solid), and accelerated upward and outward by the pressure of the wave, reach a speed in the moment that raise from the ground, that is function of the distance from the impact point, and some parameters of the wave (D and $\Delta r$).

Obtaining the derivative with respect to r, of the equation (9) and since E is constant, we obtain, after some algebra an expression for v (see Appendix, theorem 5):

$$v = v_0 \frac{rg_0}{r_0 g} \exp\left[-\int_{r_0}^{r} \frac{dr}{2Dg\Delta r}\right] \quad (10)$$

Where $r_0$ is a constant, $v_0$ is the speed of the ejecta at $r = r_0$, D is the dimensionless constant of theorem 3 (see appendix) and $\Delta r$ is the thickness of the wave.

From equation (6) we observe that when the wave reaches the radius R (final radius of the crater) it has, still, pressure (i.e., energy) but material is no longer ejected. Then at $r = R$, $Dg = \infty$ because $\rho v^2 = 0$. As D is a constant then $g = \infty$. Then g grows with r, reaching $\infty$ at $r = R$

At low r, most of the wave energy converts in kinetic, then at $r \approx 0$, $Dg \approx 1$.

One function that satisfy these conditions is:

$$g = \frac{R}{D(R-r)} \quad (11)$$

Substituting equation (11) in equation (10) we obtain (See appendix, theorem 6):

$$v = v_0 \frac{r(R-r)}{r_0(R-r_0)} \exp\left[-\frac{r-r_0}{2\Delta r} + \frac{1}{4}\frac{r^2 - r_0^2}{R\Delta r}\right] \quad (12)$$

For some cases a dimensionless form of equation (12) it is useful (see Appendix, theorem 7):

$$v^* = v_0^* \frac{r^*(1-r^*)}{r_0^*(1-r_0^*)} \exp\left[-\frac{r^* - r_0^*}{2D}\frac{R}{\Delta r} + \frac{1}{4}\frac{r^{*2} - r_0^{*2}}{D}\frac{R}{\Delta r}\right] \quad (13)$$

3. COMPARISON WITH OBSERVATIONS

We made comparisons with experimental data to validate the model as follows. In figure 2 we can see some of these comparisons.

In order to obtain subsonic impact velocity data, we make an experiment, in which steel and glass spheres were impacted, at different velocities (< 0.1 km/s), on a sand surface. Our experiments consisted of impacts classified as very low velocity with four glass projectiles (diameters of 1.23, 1.65, 2.11 and 2.42 cm) and four of steel (diameters of 1.27, 1.58, 2.21 and 2.53 cm) on a granular surface classified as sand very coarse according to the Wentworth granulometric scale (sand that passes through the N10 (2 mm) sieve and retained on the N18 (1 mm) sieve).

The experimental arrangement was configured as a containment structure to avoid sand diffusion in the laboratory, an air cannon, a glass tray (to contain the material to be hit), a high-speed camera and the spheres (projectiles). The assembly and recordings were carried out in the Hydrodynamics and Turbulence Laboratory of the Faculty of Sciences of the National Autonomous University of Mexico, in the campus of University City.

Each of the eight spheres (glass and steel) were propelled onto the target (sand) with four different pressures of the air cannon chamber (10, 20, 30 and 40 psi), obtaining a total of 32 hits. The impact velocity for each of the spheres was determined through the recordings, scaling the images (frames) with the diameter of the sphere and measuring the time between frames, all velocities was < 0.1 km/s. Recordings were made at 3,200 frames per second and the camera was placed in the horizontal plane.

From each impact recording, the ballistic trajectories of various particles (15 per axis), which are part of the ejection curtain, were measured. The trajectories were measured from the point of impact to the point that was the rim of the crater. Once the trajectory of each particle was measured frame by frame, the exit velocity and the exit angle were determined. The ejection velocity v of the ejecta was measured for different distances r*. The complete description of this experiment see Suarez-Cortes et al. (2021).
In figures 1A and 1B we can see two trials of our experiment, compared with our model.

We obtain supersonic data from literature (Anderson et al. 2003, 2007; Tsujido et al. 2015). Tsujido et al. obtained their data from Housen and Holesapple (2011) and

Cintala et al. (1999). In the experiments of Anderson et al., the impact velocity is of the order of 1 km/s (figure 2C and 2D). And in Tsujido et al. data the impact velocity are 1.7 km/s and 1.9 km/s respectively (figure 2E and 2F).

The parameters that adjust the model in each case are:
Figure 2A: $v_0^* = 1.6$; $r_0^* = 0.03$; $D = 3$; $\Delta r = 0.01$; $R = 8.34 \times 10^{-2}$ m.
Figure 2B: $v_0^* = 25$; $r_0^* = 0.03$; $D = 15$; $\Delta r = 0.01$; $R = 8.34 \times 10^{-2}$ m.
Figure 2C: $v_0^* = 500$ m/s; $r_0^* = 0.00318$ m; $D = 70$; $\Delta r = 0.001$ m; $R = 8.1 \times 10^{-2}$ m.
Figure 2D: $v_0 = 100$ m/s; $r_0 = 0.00318$ m; $D = 15$; $\Delta r = 0.002$ m
Figure 2E: $v_0^* = 160$; $r_0^* = 0.05$; $D = 5$; $\Delta r = 0.001$ m; $R = 3 \times 10^{-2}$ m.
Figure 2F: $v_0^* = 200$; $r_0^* = 0.05$; $D = 5$; $\Delta r = 0.001$ m; $R = 3 \times 10^{-2}$ m.
We take impactor radii as $r_0$.

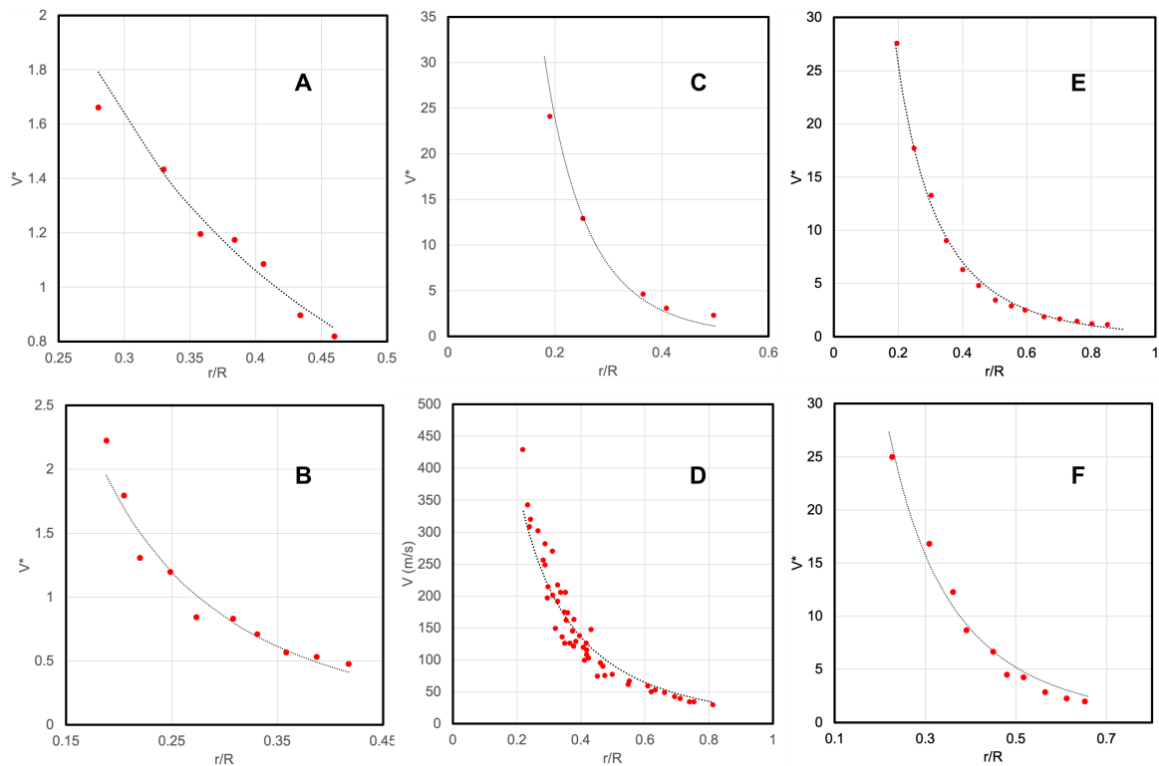

Figure 2. Validation of the model comparing with experimental data. Dotted curves are the model, red points are the data. A and B are data from our experiment. C is the data of Anderson et al. (2003). D is the data of Anderson et al. (2007), E and F are data of Tsujido et al. (2015).

## 4. DISCUSSION
Our model does not consider impact angle and assume large impact angles. We cannot compare with different impact angle experiments because we cannot find these one at literature. The velocities given by Anderson and Schultz (2006) are velocities already at ballistic fly no at beginning of this.

As we said before, the devil is in the details, and as the predictions of the model fits very well with experimental data (see figure 2) we can affirm that our model can avoid the details and deal with safety with the complexities of the process.

One problem with observations that we use to validate the model is that were all conducted on cohesionless sand targets but experiments into other materials different than sand, perhaps solid rock, we could not find at literature. Do not misunderstand us; Rock impact experiments are in the literature, but there are no ejection velocity measurements that we can compare with the theoretical results.

The most relevant result of this study is the equation (12) (Theorem 6). This result shows that ejecta speed, in our model, is only dependent of the distance from impact point and two parameters of the wave. Characteristics of the impactor and the target do not take part as independent variables of v. Of course, $v_0$ and the characteristic of the wave depends implicitly on the energy of impactor and the properties of the target, but they are not explicit in the model. This feature **of** equation (12) allows to use it at subsonic, supersonic and maybe hypersonic impacts. It is clear from figure 2 that the model works for subsonic and supersonic impacts. The question if the model also describes the ejecta behavior in hypersonic impacts ($v_i$ > 5 km/s) is still open. However, as the model derivation no depends on the impactor velocity it is likely that also work with hypersonic speed impacts. It is important to remark that differences between subsonic and supersonic impact velocities in our data is huge and even so, the model works at both cases.

We can see that velocity in the impact point (r = 0) at equation (12) is undefined, then the model only works between the range of values $0 < r_0 < r \leq R$. The value of $r_0$ is not restricted, for this very reason we choose it, in our numeric calculations, as the impactor radius, but this is not mandatory.

In the model we consider that thermal dissipation is negligible compared to kinetic energy of ejecta and energy of the wave. Of course, when the wave ends excavating and propagates beyond the crater radius, i.e., after the excavation process, all its remainder energy dissipates, but this is outside the scope of the model. At hypersonic case it is likely that model needs some adjusts because in that case thermal dissipation can be significant.

Equation (12) can be expressed as the dimensionless equation (13) (theorem 7). In fact, the equation (13) is used in graphs A, B, C, E, and F and for graph D it is used equation (12), this is because Anderson et al. (2007) do not mention crater radius value in their experiment, which is necessary to obtain dimensionless velocity.

From equation 6 (theorem 3) we obtain that the wave pressure is Dg times the ejecta kinetic energy density. From theorem 4 we know that Dg > 1, i.e., for the ejecta could be thrown, pressure could not be lesser than the density of kinetic energy in ejecta, statement that is support by equation (7) (theorem 2). Thus, while crater excavation was occurring, pressure is always larger than ejecta kinetic density. If $1 \geq$ Dg the crater formation it is impossible.

Throughout the paper we consider an impact angle close to vertical, because the experimental data of comparison are vertical shots, but the model can describe more small angles only putting v0 and D as function of $\varphi$, the azimuthal angle. As can be seen in Anderson and Schultz (2006) velocity of the ejecta at low angles is the greatest in the direction of the flight of the impactor and the smallest in the opposite. Thus, we can use, for describe these variations, cyclic functions as:

$$v_0(\varphi) = C_1 sen\left(\frac{\varphi}{2}\right) + C_2$$

$$D(\varphi) = C_3 sen\left(\frac{\varphi}{2}\right) + C_4$$

With $C_1$, $C_2$, $C_3$, and $C_4$ constants and $\varphi = 0$ in the opposite direction of impactor movement. We will work in this direction.

5 CONCLUSIONS

Our model, starting from the first principle of thermodynamics, can describes the speed of the ejecta recursing to considerations that include complex process in simple calculations.

The fit of the model with observations shows that the many complex details implicit in an impact process could be included in some few parameters (D, $v_0$, $r_0$, $\Delta r$).

Ejecta speed could be described independent of impactor parameters.

ACKNOWLEDGMENTS

Authors would like to thanks David Porta Zepeda, Carlos Echeverria Arjonilla, and all the staff from the Hydrodynamics and Turbulence Laboratory for having provided their facilities and equipment for carrying out our experiment.

APPENDIX

In this appendix we present and demonstrate the main equations that we use throughout this work. We separate them in seven theorems.

A1 DEFINITION OF THE SYSTEM

Our study has as its main objective the estimation of the speed at which the ejecta is expelled from the impacted surface. To develop an adequate formalism, we define the system of study as a hemisphere of the target's material (see Figure 1) with the next properties:
a) During the impact process, this material is displaced to form a crater.
b) The point of impact, O, is in the center of the hemisphere.
c) The radius of the hemisphere, R, is equal to the final transient crater radius.
d) The density of the system material, $\rho$, is constant and equal in all points.
e) T is the time when the crater arrives to its radius R.
f) At time t after the impact, t < T, the hemisphere excavated has radius r < R.

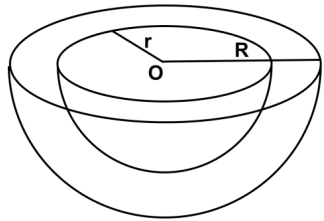

Figure 1. Diagram of the system

A2 SOME ASSUMPTIONS OF THE MODEL
a) The energy involved in the process proceed of one unique source, the projectile.
b) At the contact, the projectile transfers all its energy to the target, where it propagates as a wave in all directions inside the target.

A2. THEOREM 1

## Hypotheses

Let H be the hemispheric fraction of **the** system, excavated, when the wave reaches a radial distance r < R from the point O, and v be the speed of the ejected material of the spherical shell of radius r.

## Thesis

The kinetic energy, $E_k$, of H(r) is:

$$E_k = \rho\pi \int_0^r v^2 r^2 dr \qquad (A1)$$

## Proof

By definition of kinetic energy density, at distance r, the kinetic energy of the exterior shell of H(r') is:

$$dE_k = \frac{1}{2}\rho v^2 dV \qquad (A2)$$

Where $\rho v^2/2$ is the kinetic energy density of target's material, and dV, is volume of the ejected shell, both at distance r of the impact place.

On the other hand, the volume of the hemispherical shell dV is:

$$dV = 2\pi r^2 dr \qquad (A3)$$

Substituting equation (A3) into equation (A2):

$$dE_k = \rho\pi v^2 r^2 dr \qquad (A4)$$

And integrating from 0 to r

$$E_k = \rho\pi \int_0^r v^2 r^2 dr \qquad (A1)$$

Q.E.D.

-----------------------------------------

### A3. THEOREM 2

Let P and v be the pressure of the wave at radius $r \leq R$ and ejecta velocity respectively at distance r. Pressure of the wave is the source of ejecta energy. Pressure, P, satisfy the inequality:

$$P > \frac{1}{2}\rho v^2 \qquad (A5)$$

## Proof

By definition, if V and $\varepsilon$ are the volume, and the kinetic energy density of a system, respectably, then total kinetic energy is:

$$E_k = \varepsilon V \qquad (A6)$$

By definition of kinetic energy density:

$$\varepsilon = \frac{1}{2}\rho v^2 \quad (A7)$$

When excavating of the crater is in process a semi-shell of radius r and thickness Δr it has a volume:

$$V = 2\pi r^2 \Delta r \quad (A8)$$

Substituting equations (A7) and (A8) in (A6):

$$E_K = 2\pi r^2 \Delta r \frac{1}{2}\rho v^2 \quad (A9)$$

Total energy in the wave is the product of volume times pressure, i.e., equation (A8) times P:

$$E_W = 2\pi r^2 \Delta r P \quad (A10)$$

By the hypothesis that P is the source of ejecta energy then not all the wave energy is converted into kinetic energy, some is transmitted with the wave, then at all moments it satisfies the next condition:

$$E_W > E_K \quad (A11)$$

From equations (A9), (A10) and (11) we have:

$$2\pi r^2 \Delta r P > 2\pi r^2 \Delta r \frac{1}{2}\rho v^2 \quad (A12)$$

And reducing:

$$P > \frac{1}{2}\rho v^2 \quad (A5)$$

Q.E.D.

----------------------------------------

A4. THEOREM 3
Hypotheses: Same as theorems 1 and 2.
Thesis: Pressure inside the wave at radius r, is Dg times the ejecta kinetic energy density at radius r, where D is a dimensionless constant and g is a dimensionless function of r:

$$P = Dg\frac{1}{2}\rho v^2 \quad (A13)$$

Proof
As we have four variables: r, distance reached by the wave, measured from the point of impact; P(r), pressure of the wave; v(r), velocity of the ejecta, and ρ, density of the impacted medium. Then we can construct a function, f, of the four variables:

$$f(P, r, v, \rho) = 0 \quad (A14)$$

The dimensions that form each variable are M, L and T. Then by the Vaschy-Buckingham $\Pi$ theorem, if we have n = 4 variables and k = 3 dimensions, therefore we have (n-k = 1) dimensionless variable, $\Pi_1$, and therefore equation (A14) can be expressed as:

$$f(\Pi_1) = 0 \quad (A15)$$

We choose r, v and ρ as basic variables and we put:

$$\Pi_1 = g(r) P r^a v^b \rho^c \quad (A16)$$

Function g(r) is a dimensionless function of r.
By the dimensionless condition

$$M^0 L^0 T^0 = M^0 L^0 T^0 M L^{-1} T^{-2} (L)^a (LT^{-1})^b (ML^{-3})^c = M^{1+c} L^{a+b-3c-1} T^{-b-2} \quad (A17)$$

Thus, solving the system for a, b, c
$$a = 0$$
$$b = -2 \quad (A18)$$
$$c = -1$$

That is:
$$\Pi_1 = \frac{g(r)P}{\rho v^2} \quad (A19)$$

Substituting equation (A19) into equation (A15):

$$f\left(\frac{g(r)P}{\rho v^2}\right) = 0 \quad (A20)$$

Making the composition of equation (A20) with its inverse relation, $f^{-1}$:

$$\frac{g(r)P}{v^2 \rho} = f^{-1}(0) = C \quad (A21)$$

Where C is a dimensionless constant.

The constant C in equation (A21) is obtained because f⁻¹, whatever it is the relation, evaluated at zero has a unique value.
Thus:
$$P = Cgv^2\rho \qquad (A22)$$

Making, $C = \frac{D}{2}$, where D is another dimensionless constant, and rearranging

$$P = Dg\frac{1}{2}\rho v^2 \qquad (A13)$$

Q.E.D.

-----------------------------------------

A5. THEOREM 4
Hypotheses: Same as theorems 1 and 2.
Thesis: The values of Dg in theorem 3 satisfy the inequality:

$$Dg > 1 \qquad (A23)$$

Proof
By theorem 2:

$$P > \frac{1}{2}\rho v^2 \qquad (A24)$$

And substituting the value of P given in theorem 3:

$$Dg\frac{1}{2}\rho v^2 > \frac{1}{2}\rho v^2 \qquad (A25)$$

And simplifying

$$Dg > 1 \qquad (A23)$$

Q.E.D.

-----------------------------------------

A6. THEOREM 5
Hypotheses: Same as theorems 1 and 2.
Thesis: The speed, v, of the ejected material at distance r from the impact point O, is:

$$v = v_0 \frac{rg_0}{r_0 g} \exp\left[-\int_{r_0}^{r} \frac{dr}{2Dg\Delta r}\right] \qquad (A26)$$

Where $r_0$ is the radius of the impactor, $v_0$ is the speed of the ejecta at $r = r_0$, D is the dimensionless constant of theorem 3, and $\Delta r$ is the thickness of the wave.

Proof

By hypothesis e) all kinetic energy (E) of the impactor is transferred to the target. And by energy conservation this energy it is distributed, in the target, as follows:

$$E = E_K + E_W + E_T \quad (A27)$$

$E_K$ is the kinetic energy of the displaced material; $E_W$ is the energy carried by the wave and $E_T$ the energy dissipated as thermal energy.
The energy carried by the wave is:

$$E_W = PV_W \quad (A28)$$

Where P is the pressure, and $V_W$ is the volume occupied by the wave.
Considering that the energy dissipated as thermal energy by the wave is small enough to neglect it, then:

$$E_T = 0 \quad (A29)$$

As the wave advances from the point of impact, O, it excavates the crater. At time t after the contact between the target and the impactor, using theorem 1 and equations (A24), (A25), and (A26), the energy E transferred to the target from the impactor, is expressed as:

$$E = \rho\pi \int_0^r v^2 r^2 dr + PV_W \quad (A30)$$

If we assume a hemispherical excavation of the crater, $V_W$ is expressed as:

$$V_W = 2\pi r^2 \Delta r \quad (A31)$$

Where $\Delta r$ is the width of the wave.
Substituting equation (A31) in equation (A30) we have:

$$E = \pi\rho \int_0^r v^2 r^2 dr + 2\pi P r^2 \Delta r \quad (A32)$$

Using theorem 2 in equation (A32):

$$E = \pi\rho \int_0^r v^2 r^2 dr + \pi D g \rho v^2 r^2 \Delta r \quad (A33)$$

Obtaining the derivative with respect to r:

$$\frac{dE}{dr} = \pi\rho v^2 r^2 + \pi D \frac{dg}{dr}\rho v^2 r^2 \Delta r + 2\pi D g \rho v \frac{dv}{dr} r^2 \Delta r + 2\pi D g \rho v^2 r \Delta r \quad (A34)$$

Multiplying by dr and as E is constant it does not change with r and therefore its differential is zero:

$$\pi\rho v^2 r^2 dr + \pi D\rho v^2 r^2 \Delta r dg + 2\pi Dg\rho v r^2 \Delta r dv + 2\pi Dg\rho v^2 r \Delta r dr = 0 \quad (A35)$$

Dividing by $2\pi\rho Dg\Delta r v^2 r^2$:

$$\frac{dr}{2Dg\Delta r} + \frac{dg}{2g} + \frac{dv}{v} + \frac{dr}{r} = 0 \quad (A36)$$

If we obtain the integral between 0 and r, the integral of the fourth term was undefined, then we need to define the lower limit of integration to values r > 0. Then integrating between $r_0 > 0$ and r:

$$\int_{r_0}^{r} \frac{dr}{2Dg\Delta r} + \ln\frac{g}{g_0} + \ln\frac{v}{v_0} - \ln\frac{r}{r_0} = 0 \quad (A37)$$

Solving for v

$$v = v_0 \frac{rg_0}{r_0 g} \exp\left[-\int_{r_0}^{r} \frac{dr}{2Dg\Delta r}\right] \quad (A26)$$

Q.E.D.

-----------------------------------------

A7. THEOREM 6
Hypotheses:
According by theorem 3:

$$Dg = \frac{2P}{\rho v^2} \quad (A38)$$

When the wave reaches the radius R (final radius of the crater) it has, still, pressure (i.e., energy) but material is no longer ejected. Then at r = R, $Dg = \infty$ because $\rho v^2 = 0$. As D is a constant then $g = \infty$. Then g grows with r, reaching ∞ at r = R
At low r, most of the wave energy converts in kinetic, then at $r \approx 0$, $Dg \approx 1$.
Let assume that g is:

$$g = \frac{R}{D(R-r)} \quad (A39)$$

Thesis

$$v = v_0 \frac{r(R-r)}{r_0(R-r_0)} \exp\left[-\frac{r-r_0}{2\Delta r} + \frac{1}{4}\frac{r^2 - r_0^2}{R\Delta r}\right] \quad (A40)$$

Proof
Substituting (A39) in (A26):

$$v = v_0 \frac{r(R-r)}{r_0(R-r_0)} \exp\left[-\int_{r_0}^{r} \frac{(R-r)dr}{2R\Delta r}\right] \quad (A41)$$

Integrating:

$$v = v_0 \frac{r(R-r)}{r_0(R-r_0)} \exp\left[-\frac{r-r_0}{2\Delta r} + \frac{1}{4}\frac{r^2 - r_0^2}{R\Delta r}\right] \quad (A40)$$

QED

Scholium
The function:

$$g = \frac{(1 - \exp(R))}{D(\exp(r) - \exp(R))}$$

Also satisfy the conditions: Dg (0) = 1 and g(R) = ∞ as (A39). Therefore, the function g is not unique.

A8 Theorem 7
Hypotheses: Same of theorem 6.
Thesis:
The expression of equation (A40) (theorem 6) in dimensionless form is:

$$v^* = v_0^* \frac{r^*(1-r^*)}{r_0^*(1-r_0^*)} \exp\left[-\frac{r^* - r_0^*}{2D}\frac{R}{\Delta r} + \frac{1}{4}\frac{r^{*2} - r_0^{*2}}{D}\frac{R}{\Delta r}\right] \quad (A42)$$

Proof
By theorem 5:

$$v = v_0 \frac{r(R-r)}{r_0(R-r_0)} \exp\left[-\frac{r-r_0}{2D\Delta r} + \frac{1}{4}\frac{r^2 - r_0^2}{DR\Delta r}\right] \quad (A43)$$

Defining $r^* = r/R$ and $v^* = v(r)/\sqrt{gR}$ where R is the radius of the transient crater and g is gravity acceleration. Substituting in equation (A43):

$$v^*\sqrt{gR} = v_0^*\sqrt{gR} \frac{r^*R(R-r^*R)}{r_0^*R(R-r_0^*R)} \exp\left[-\frac{r^*R - r_0^*R}{2D\Delta r} + \frac{1}{4}\frac{r^{*2}R^2 - r_0^{*2}R^2}{DR\Delta r}\right] \quad (A44)$$

Simplifying:

$$v^* = v_0^* \frac{r^*(1-r^*)}{r_0^*(1-r_0^*)} \exp\left[-\frac{r^* - r_0^*}{2D}\frac{R}{\Delta r} + \frac{1}{4}\frac{r^{*2} - r_0^{*2}}{D}\frac{R}{\Delta r}\right] \quad (A42)$$

QED